\documentclass[aps,prb,twocolumn,superscriptaddress, showpacs,floatfix]{revtex4}
\usepackage{graphicx,epsfig,amsmath}

\newcommand{\subfig}[2]{Fig.~\ref{fig:#1}(#2)} 

\newcommand{\Vds}{\mbox{$\text{V}_{\text{ds}}$}}
\newcommand{\VQPC}{\mbox{$\text{V}_{\text{QPC}}$}}

\newcommand{\ETQD}{\mbox{$\text{E}_{\text{TQD}}$}}
\newcommand{\C}[2]{C$_{\text{#1},\text{#2}}$}
\newcommand{\NL}{\mbox{$\text{N}_{\text{L}}$}}
\newcommand{\NC}{\mbox{$\text{N}_{\text{C}}$}}
\newcommand{\NR}{\mbox{$\text{N}_{\text{R}}$}}

\begin{document}

\title{The 3D transport diagram of a triple quantum dot}


\author{G.~Granger}
	\email{Ghislain.Granger@nrc.ca}
	\affiliation{Institute for Microstructural Sciences, National Research Council Canada, Ottawa, ON Canada K1A 0R6}
\author{L. Gaudreau}
	\affiliation{Institute for Microstructural Sciences, National Research Council Canada, Ottawa, ON Canada K1A 0R6}
	\affiliation{D\'epartement de physique, Universit\'e de Sherbrooke, Sherbrooke, QC Canada J1K 2R1}
\author{A.~Kam}
	\affiliation{Institute for Microstructural Sciences, National Research Council Canada, Ottawa, ON Canada K1A 0R6}
\author{M. Pioro-Ladri\`ere}
	\affiliation{D\'epartement de physique, Universit\'e de Sherbrooke, Sherbrooke, QC Canada J1K 2R1}
\author{S.A.~Studenikin}
	\affiliation{Institute for Microstructural Sciences, National Research Council Canada, Ottawa, ON Canada K1A 0R6}
\author{Z.R. Wasilewski}
	\affiliation{Institute for Microstructural Sciences, National Research Council Canada, Ottawa, ON Canada K1A 0R6}
\author{P. Zawadzki}
	\affiliation{Institute for Microstructural Sciences, National Research Council Canada, Ottawa, ON Canada K1A 0R6}
\author{A.S.~Sachrajda}
	\affiliation{Institute for Microstructural Sciences, National Research Council Canada, Ottawa, ON Canada K1A 0R6}


\begin{abstract}

We measure a triple quantum dot in the regime where three addition lines, corresponding to the addition of an electron to each of three dots, pass through each other. In particular, we probe the interplay between transport and the tridimensional nature of the stability diagram. We choose the regime most pertinent for spin qubit applications. We find that at low bias transport through the triple quantum dot circuit is only possible at six quadruple point locations. The results are consistent with an equivalent circuit model.

\end{abstract}


\maketitle


\section{Introduction}

Double quantum dots have been extensively studied and utilized in quantum information spin qubit experiments.\cite{Hanson2007, PioroLadriere2008, Bluhm2010} Recently we have presented a highly tunable triple quantum dot (TQD) layout \cite{Gaudreau2009} aimed at quantum information applications, where the stability diagram was mapped out using standard charge detection techniques. The experimental realization of this type of tunable TQD is important, as it provides a platform for testing a variety of predicted novel quantum information functionalities, running simple algorithms, and investigating interference effects. \cite{Loss1998, Tanamoto2000, DiVincenzo2000, Saraga2003, Greentree2004, Fabian2005, Michaelis2006, Busl2010, Vernek2009}

Electron transport in double quantum dot systems is limited to the neighborhood of special locations in the stability diagram where three electronic configurations are degenerate (triple points).\cite{Hanson2007} In this paper we probe the equivalent transport conditions in TQDs. In the region of a stability diagram where one electron is added to each of the three dots, it has been predicted that electrostatics would result in a total of six quadruple points (QPs).\cite{Rogge2009} While the stability diagram of a TQD circuit has been studied in detail using conventional charge detection techniques \cite{Gaudreau2006, Rogge2009, Schroer2007, Gaudreau2009} and certain transport features have been observed in these experiments,\cite{GaudreauPRB2009, Rogge2009, Schroer2007} in this paper we reveal the full interplay between the stability diagram and the conditions for electron transport. 

\begin{figure}[tbh]
\setlength{\unitlength}{1cm}
\begin{center}
\begin{picture}(8,6.7)(0,0)
\includegraphics[width=8cm, keepaspectratio=true]{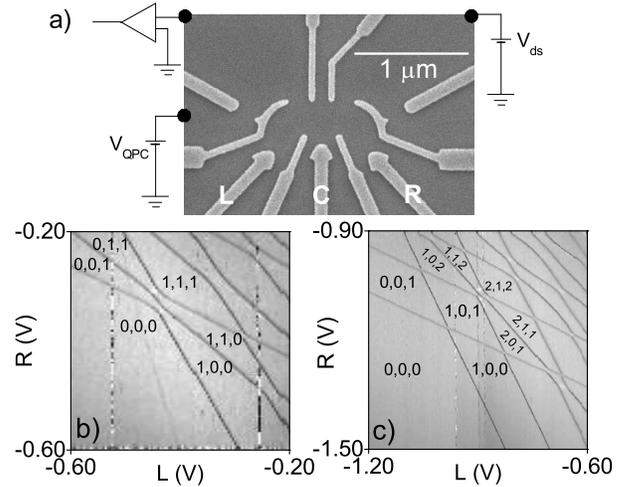}
\end{picture}
\end{center}
\caption{(a) Electron micrograph of a device similar to the ones measured. The plunger gate for the left, center, and right dots are labeled L, C, and R, respectively. Black dots represent Ohmic contacts. \Vds~is the drain-source bias across the TQD, and \VQPC~is the bias across the QPC. The current preamplifier is shown schematically. (b) and (c) Charge detection transconductance measurements in the L-R space, at \Vds=0~mV plotted with a greyscale where black is low and white is high. (b) The absolute electronic configurations from (0,0,0) to (1,1,1) are indicated. \VQPC=0.3~mV (c) The electronic configurations from (0,0,0) to (2,1,2) are indicated.}
\label{fig:1}
\end{figure}

\begin{figure*}[bth]
\setlength{\unitlength}{1cm}
\begin{center}
\begin{minipage}[c]{0.72\linewidth}
\includegraphics[width=12.8cm, keepaspectratio=true]{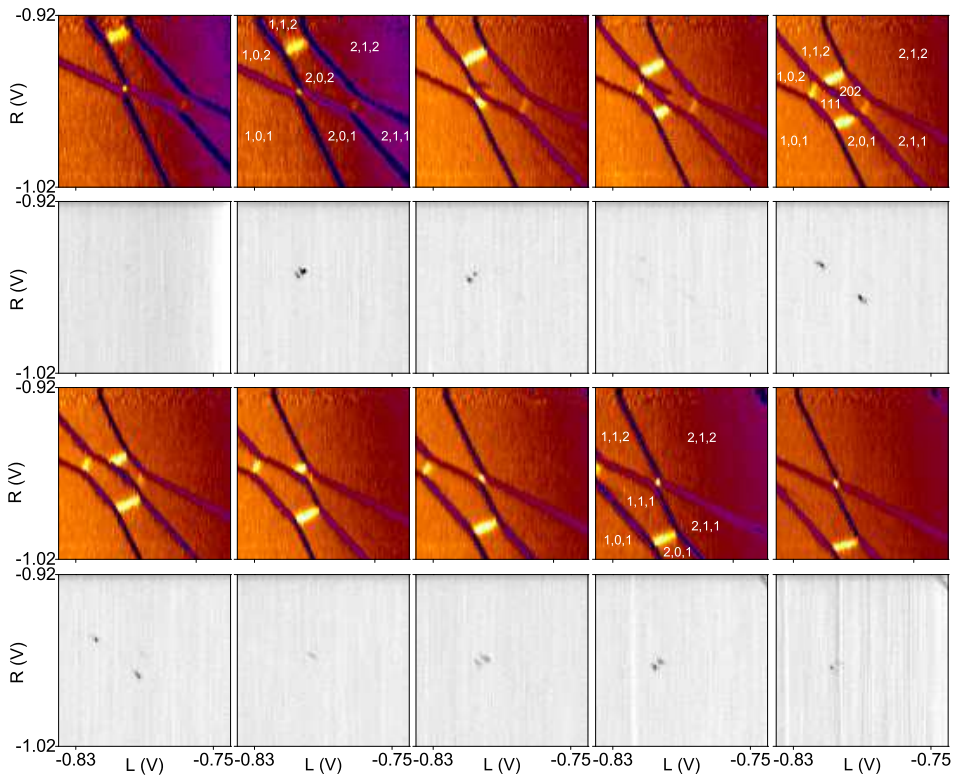}
\end{minipage}\hfill
\begin{minipage}[c]{0.25\linewidth}
\caption{(Color online.) Charge detection and transport diagrams in the L-R voltage space as a function of gate voltage C. The values of C in the upper two rows go from -0.224~V to -0.216~V, while they go from -0.214~V to -0.206~V in the lower two rows. The charge detection transconductance data at \Vds=0~mV (1$^{\text{st}}$ and 3$^{\text{rd}}$ rows) are plotted with an arbitrary unit colorscale: dark blue (black) is low transconductance, orange (grey) is medium, and yellow (white) is high. The transport data (2$^{\text{nd}}$ and 4$^{\text{th}}$ rows) are taken with \Vds=0.1~mV across the TQD. The greyscale from white to black corresponds to a current range of 370~fA through the TQD. The electronic configurations are indicated on the charge detection plots for C={-0.222, -0.216, -0.208}~V, where clear black spots in the respective transport plots are observed.}
\label{fig:2}
\end{minipage}
\end{center}
\end{figure*}

We first use a charge detector \cite{Field1993} to set up specific electronic configurations. We focus mainly on the regime between (1,0,1) and (2,1,2), which is the most relevant for spin qubit applications. The 3D nature of the stability diagram and transport regimes is studied by tuning plunger gate voltages which control the occupation numbers of the three individual dots. We compare our results to an equivalent circuit model.\cite{Gaudreau2006, Schroer2007, Vidan2004}

\section{Experimental details}

The devices are fabricated on a chip from a GaAs/AlGaAs heterostructure grown by Molecular Beam Epitaxy with a density of 2.1$\times10^{11}$~cm$^{-2}$ and a mobility of 1.72$\times10^6$~cm$^2$/Vs. Annealed NiAuGe Ohmic contacts are used to contact the two-dimensional electron gas (2DEG) located 110~nm below the surface of the wafer. TiAu gate electrodes are patterned by electron-beam lithography to allow electrostatic control of the triple dot potential, as shown in \subfig{1}{a}. Two gates are used to define charge detectors on the left and right of the TQD, but only the left quantum point contact (QPC) is used for the data shown in this paper.

Both charge detection and transport measurements are made in the voltage plane determined by the outer plunger gates L and R (see micrograph in \subfig{1}{a}), which are located close to the left and right dots, respectively. In the case of charge detection \cite{Field1993}, the left QPC transconductance at \Vds=0 is measured with a lock-in technique using a typical root-mean-square modulation of 1~mV on gate R (\subfig{1}{a}). The QPC detector conductance is tuned to a high sensitivity point (near e$^2$/h) and the bias across it is \VQPC=0.2~mV unless indicated otherwise. In the case of transport, a drain-source bias \Vds=0.1~mV is applied on the upper right hand side Ohmic contact (\subfig{1}{a}), and the DC current is measured with a current preamplifier attached to the upper left hand side Ohmic contact. 

Samples are bias-cooled with 0.25~V on all gates in a dilution refrigerator. The electron temperature is $\approx$110~mK. We probe the tridimensional nature of the stability diagram by tuning the center gate plunger voltage, C, for fixed sweeps of plunger gates L and R, which allows us to move anywhere in the L-R-C tridimensional gate voltage space.  The remaining gates are biased negatively in such a way to electrostatically form the appropriate potential for the triple dot.

\section{Results}

Figures \ref{fig:1}(b) and (c) contain transconductance data from two TQD devices in the few-electron regime. The relevant regime for charge qubits \cite{Tanamoto2000} is shown in \subfig{1}{b} and that for spin qubits,\cite{Loss1998, Laird2010} in \subfig{1}{c}. These electronic configurations can be achieved in both the devices that have been measured (not shown). 

Figure~\ref{fig:2} shows the evolution of the 3D stability diagram as a function of C, the voltage on the central plunger gate. The electronic configurations (N$_{\text{L}}$,N$_{\text{C}}$,N$_{\text{R}}$) are identified by counting from the (0,0,0) configuration shown in \subfig{1}{c}. 

The region of interest in Fig.~\ref{fig:2} is the regime where three addition lines cross between (1,0,1) and (2,1,2). Depending on the value of C, the size of the stability diagram regions with electronic configurations (1,1,1) and (2,0,2) can be tuned continuously from no (1,1,1) region and a large (2,0,2) region to a large (1,1,1) region and no (2,0,2) region. A symmetric case where the (1,1,1) and (2,0,2) regions are approximately the same size is also observed. The latter regime is important when considering transport.

The bright lines in Fig.~\ref{fig:2} correspond to charge transfer lines between dots that do not involve any changes in the total number of electrons. The boundary between (1,1,1) and (2,0,2) is an example of a charge reconfiguration (commonly referred to as a QCA event due to its similarity to Quantum Cellular Automata \cite{Gaudreau2006}) at which every dot changes its electron occupation as the total number of electrons changes by one.  The system can minimize its energy by spreading out the charge for the three electrons in (1,1,1), but for four electrons the system minimizes its energy by going to separated pairs of electrons in (2,0,2). The QCA effect is unique to TQD systems; it does not occur in double dots.

We now turn to examining the transport through the TQD. The DC current measurements in Fig.~\ref{fig:2} are shown as a function of the C gate voltage, and each transport diagram can be compared to the charge detection stability diagram in the row immediately above it. As the voltage on C progressively increases, two closely spaced current spots first appear and disappear, then two well separated current spots appear and disappear, and two additional closely spaced current spots appear and disappear. The total number of current spots is six. 

The simultaneous appearance of pairs of current spots is interesting. We focus on the first two current spots from Fig.~\ref{fig:2}, which we label QPs 1 and 2. Figure~\ref{fig:3} contains their C gate dependence.\cite{footnote2} The data in Fig.~\ref{fig:3} are taken with a slightly higher resolution than in Fig.~\ref{fig:2}, so the current spots now appear as roughly triangular regions.  As C is made less negative, the current intensity of the two triangles changes so that QP 2 gets progressively replaced by QP~1.

For clarity we reproduce in Fig.~\ref{fig:4} the charge detection and transport data from Fig.~\ref{fig:2} where clear pairs of transport spots occur. It can be seen that transport occurs only at QPs. The six QPs are labeled from 1 to 6. The inset in the left panel confirms that the transport spots occur as the (1,1,1) region closes. Similarly, the inset in the right panel, taken at a nearby C voltage, indicates that the transport spots occur as the (2,0,2) region closes. Table~\ref{tab:1} lists the four degenerate electronic configurations at each QP. 

The curved arrows in Fig.~\ref{fig:4} indicate single-particle sequences involving only nearest neighbor events for transport (in analogy with the usual triple point sequential transport sequences in double quantum dots). Thus, for QP~1, an electron can be transferred through the device by the following sequence. Starting in (1,0,1), an electron enters the left dot and transfers from dot to dot, exiting the right dot. Such a process requires all four configurations (1,0,1), (2,0,1), (1,1,1), and (1,0,2). In a similar fashion, for QP~6, hole transport occurs from right to left by going through the sequence (2,1,2), (2,1,1), (2,0,2), and (1,1,2).\cite{footnote3}

For QPs~2 and 5, the equivalent sequences are slightly more subtle. The sequence for QP~2, for example, starts with (1,0,2). An electron is added to the left dot from the left lead to reach (2,0,2). Then the electron from the right dot escapes to the right lead to give (2,0,1), which gives rise to a net current. The system returns to the initial configuration via the (1,1,1) state. Note that the current carrying sequences for QPs 1, 2, 5, and 6 all correspond to a simple closed curve passing through the relevant four degenerate configurations similar to that at triple points in double quantum dots. These curves are drawn in the left and right panels in Fig.~\ref{fig:4}. 

In contrast, the sequences for QPs 3 and 4 differ from all the others, as their trajectories describe figures of eight, shown in the central panel of Fig.~\ref{fig:4}. Such figures of eight do not have an analog in double dots. For QP~3, we start with (1,0,2). An electron hops into the left dot from the left lead to reach (2,0,2). Then, a charge transfer occurs between the left and center dots to reach (1,1,2), and an electron from the right dot hops off to the right lead to get (1,1,1) and produce net current flow. Finally, a charge transfer from the center dot to the right dot occurs to restore the (1,0,2) configuration. Likewise, a similar process occurs for QP~4.

\section{Analysis}

\begin{figure}[bth]
\setlength{\unitlength}{1cm}
\begin{center}
\begin{picture}(8,4)(0,0)
\includegraphics[width=8.0cm, keepaspectratio=true]{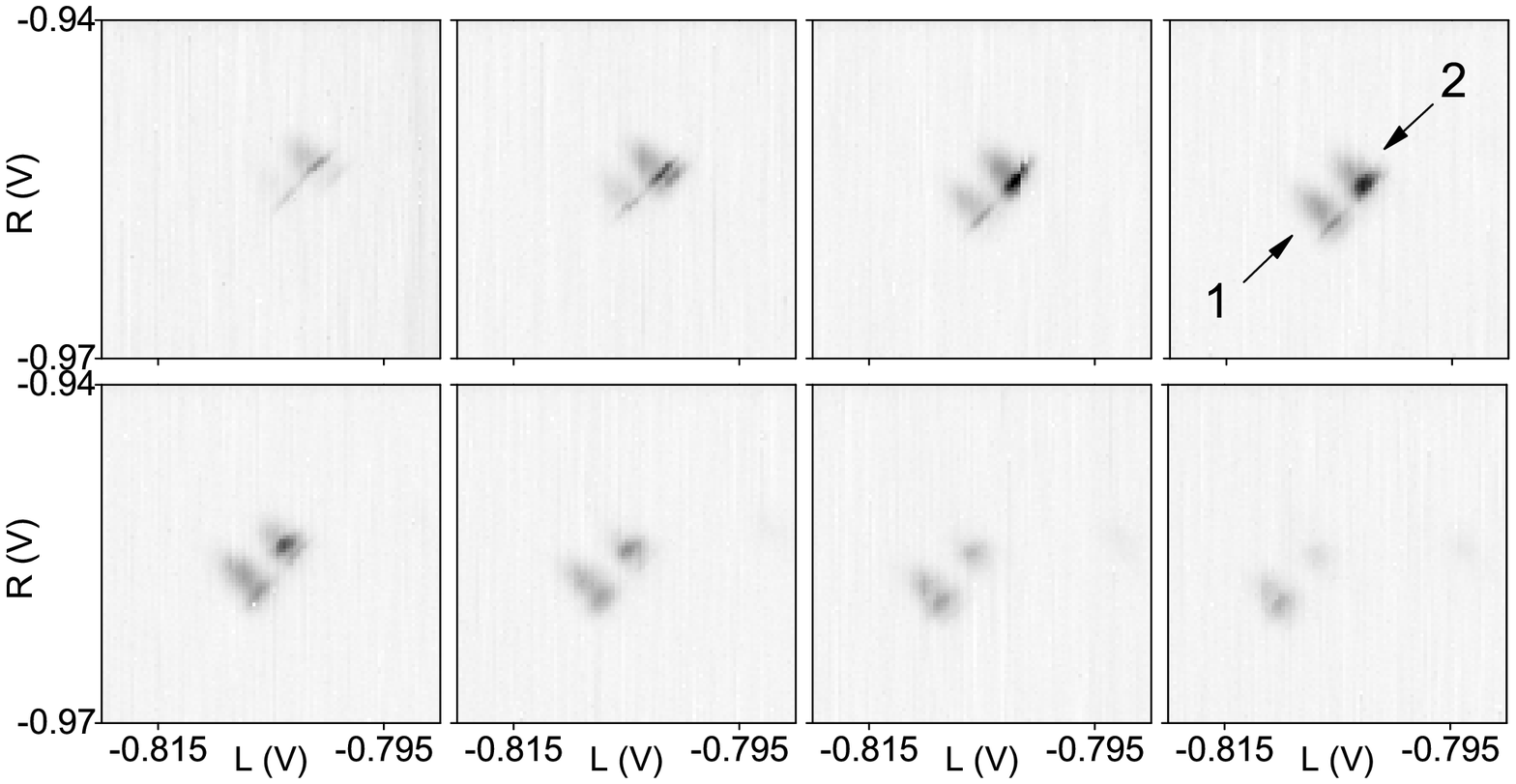}
\end{picture}
\end{center}
\caption{C gate voltage dependence of DC current through the TQD in the R-L voltage plane near QPs 1 and 2 at \Vds=0.1~mV. C goes from -0.218~V to -0.2165~V in the 1$^{\text{st}}$ row and from -0.216~V to -0.2145~V in the 2$^{\text{nd}}$ row. The greyscale from white to black corresponds to a current range of 730~fA.}
\label{fig:3}
\end{figure}

\begin{figure}[tbh]
\setlength{\unitlength}{1cm}
\begin{center}
\begin{picture}(8,5.5)(0,0)
\includegraphics[width=8cm, keepaspectratio=true]{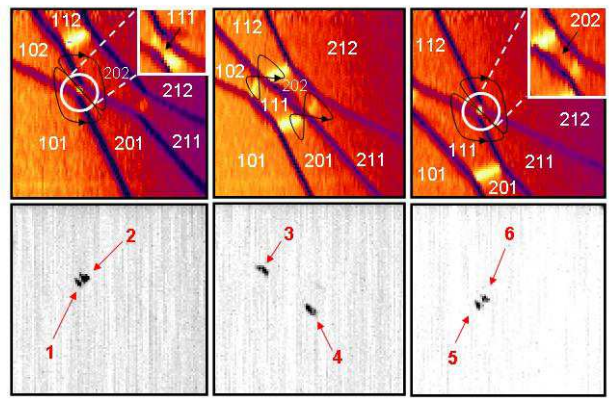}
\end{picture}
\end{center}
\caption{(Color online.) Charge detection transconductance and DC transport reproduced from Fig.~\ref{fig:2}. The values of C in the left, middle, and right panels are -0.222~V, -0.216~V, and -0.208 V, respectively. The charge detection transconductance data are plotted with an arbitrary unit colorscale: dark blue (black) is low, orange (grey) is medium, and yellow (white) is high. The electronic configurations are indicated. The insets in the left and right panels are from C=-0.220~V and -0.214~V, respectively. Arrows indicate sequences for current flow.}
\label{fig:4}
\end{figure}

In order to gain further understanding of the C gate evolution, we use the equivalent circuit model \cite{Gaudreau2006, Schroer2007, Vidan2004} to identify where QPs occur within stability diagram. This model allows us to first calculate the energy of the TQD, \ETQD, for each of the eight relevant electronic configuration between (1,0,1) and (2,1,2) everywhere in the 3D voltage space. The expression for \ETQD~is found in Eq.~(\ref{eqn:1}), where $N_i$ is the number of electrons on dot $i$, $U_{i,j}$ is the charging energy if $i=j$ or the capacitive coupling energy between dots if $i \neq j$, and $f$ gives the electrostatic energy originating from the gate voltages $\vec V$ with a capacitance matrix C between the gates and the dots in the $\vec N$ electronic configuration.

\begin{equation}
E_{TQD}(\vec N, \vec V)=\sum_{i,j=1}^{3}\frac{1}{2}N_iN_jU_{i,j}+ f(\vec N, \vec V, C)
\label{eqn:1} 
\end{equation}

To perform the calculations, we make assumptions about the relative sizes of the single electron charging energies, the mutual capacitive energies for electrons on different dots, and the gate-dot capacitances \C{gate}{dot}. The mutual capacitive energy between adjacent (outer) dots  is taken as 0.1 (0.01) times the charging energy. \C{gate}{dot} are taken such that  \C{L}{C}=0.6\C{L}{L}, \C{L}{R}=0.4\C{L}{L}, and the rest of the capacitance matrix follows by symmetry.

\begin{table}[htb]
\begin{center}
		\begin{tabular}{|c|c|c|}

\hline
QP &Configurations & $n\times N$	\\ \hline
1  & (1,0,1)	(1,1,1)	(1,0,2)	(2,0,1) &	1$\times$2 	and		3$\times$3 	\\ \hline
2  & (1,1,1)	(1,0,2)	(2,0,1)	(2,0,2) &	3$\times$3 	and		1$\times$4 	\\ \hline
3  & (1,1,1)	(1,1,2)	(1,0,2)	(2,0,2) &	2$\times$3 	and		2$\times$4 	\\ \hline
4  & (1,1,1)	(2,0,1)	(2,1,1)	(2,0,2) &	2$\times$3	and		2$\times$4	\\ \hline
5  & (1,1,1)	(1,1,2)	(2,1,1)	(2,0,2) &	1$\times$3	and		3$\times$4 	\\ \hline
6  & (1,1,2)	(2,1,1)	(2,0,2)	(2,1,2) &	3$\times$4	and		1$\times$5	\\ \hline

		\end{tabular}
		\caption{Degenerate electronic configurations for the six QPs when adding one electron to the three quantum dots from (1,0,1) to (2,1,2). The number of configurations $n$ with a given total electron number $N$ are also indicated as $n\times N$.}
\label{tab:1}
\end{center}
\end{table}

\begin{figure}[htb]
\setlength{\unitlength}{1cm}
\begin{center}
\begin{picture}(8,5.5)(0,0)
\includegraphics[width=8.0cm, keepaspectratio=true]{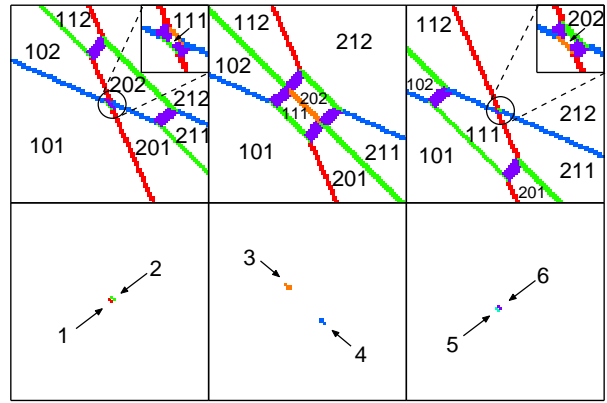}
\end{picture}
\end{center}
\caption{(Color online.) 2D cuts from the calculated 3D stability diagram and QPs with increasing center plunger gate voltage from left to right. (top row) Stability diagrams, where red (medium grey), green (lighter grey), and blue (dark grey) thin lines correspond to adding an electron to the left, center, and right dot, respectively;  purple (thick grey) lines correspond to charge transfers; and the orange (lightest grey) line corresponds to the QCA effect. Insets are zooms at values of C slightly off from the QP. (bottom row) The six QPs found from the model are labeled from 1 to 6.}
\label{fig:5}
\end{figure}

In order to visualize the 3D stability diagram for the 8 electronic configurations between (1,0,1) and (2,1,2), one needs to find the location of several planes; namely, the planes where an electron is added to left dot; the planes where an electron is added to the right dot; the planes where an electron is added to the center dot; the planes where an electron is transferred between two dots; and the plane where the QCA effect occurs between the (1,1,1) and (2,0,2) configurations. The latter plane is a mixed plane where the addition of an electron and the charge transfer occur simultaneously.

All the planes that involve a change in the total number of electrons (i.e.~including the QCA plane) are found by locating points in the 3D voltage space where \ETQD(\NL,\NC,\NR) is degenerate for two configurations (also requiring the two configurations to be the lowest energy states of the system). For the charge transfer planes, we require that the electrochemical potential $\mu$(\NL,\NC,\NR) of two configurations be equal (also requiring that the two configurations be the lowest energy states of the system). We define $\mu$(\NL,\NC,\NR)=\ETQD(\NL,\NC,\NR)-\ETQD(\NL+\NC+\NR-1), where the last term is evaluated in the ground state of the \NL+\NC+\NR-1 configuration subspace. The calculated 3D stability diagrams are shown in the top row of Fig.~\ref{fig:5}.

The QPs are part of the stability diagram, as, by definition, they occur wherever the value of \ETQD~is the same for four different electronic configurations. We use the equivalent circuit model to find where this four-fold degeneracy occurs. The results are drawn in the bottom row of Fig.~\ref{fig:5} for the same C gate voltages as in the top row. The six QPs revealed numerically for the crossing of three addition planes are comparable to the measurements of Figs.~\ref{fig:2} to ~\ref{fig:4}, and they occur in three subsequent pairs as the C gate voltage is made more positive. This is different from the transport at triple points in double dots, which occurs independently of the gate voltage on a third gate.

\section{Conclusion}

In conclusion, we have studied the interplay between transport and the 3D stability diagram of a triple quantum dot circuit in the regime with the electron configurations between (1,0,1) and (2,1,2), relevant for spin qubits. We have found that transport occurs only at each of six quadruple points. The six quadruple points are consistent with a simple equivalent circuit model. Understanding of this interplay is important for quantum information experiments requiring transport through triple quantum dot circuits.

\acknowledgments

We thank D.G. Austing, P. Hawrylak, and C.Y. Hsieh for discussions. A.S.S. acknowledges funding from NSERC Grant No. 170844-05. A.S.S. and A.K. acknowledge funding from CIFAR. G.G. acknowledges funding from the NRC-CNRS collaboration. M.P.-L. acknowledges funding from NSERC and CIFAR.


\begin{thebibliography}{00}
\bibitem{footnote2} The exact values of C are slightly less negative here compared to the data in Figs.~\ref{fig:2} and \ref{fig:4} because of a background charge jump between the experiments.
\bibitem{footnote3} Equivalently, one can think about transport of an electron at QP~6 from left to right through a cyclic permutation of the hole transport sequence starting with (1,1,2).
\end{thebibliography}
\end{document}